\documentstyle[12pt]{article}
\def\>{\rangle}
\def\<{\langle}
\def\|{|\!|}
\def\H{{\cal H}}
\def\ir{{\rm I}\hskip-.2em{\rm R}}
\def\{{\lbrace}
\def\}{\rbrace}
\def\no{\nonumber}
\def\E{{\rm E}\hskip-.53em{\rm I}\;}
\def\half{{\textstyle{1\over2}}} 
\def\o{\overline}

\def\D{{\cal D}}

\def\one{1\hskip-.37em 1}
\def\ra{\rightarrow}

\def\tint{{\textstyle{\int}}}
\def\bea{\begin{eqnarray}}
\def\eea{\end{eqnarray}}
\def\P{{\cal P}}

\def\>{\rangle}
\def\<{\langle}
\def\|{|\!|}
\def\H{{\cal H}}
\def\half{{\textstyle{1\over2}}} 
\def\quarter{{\textstyle{1\over4}}} 
\def\o{\overline}

\def\one{1\hskip-.37em 1}
\def\d{\partial}

\title{Coherent State Path Integrals \\ at (Nearly) 40}
\date{}
\author{John R.~ Klauder\\
Department of Physics \footnote{Also Department of Mathematics.}, 
University of Florida\\
 P.O. Box 118440, Gainesville, FL 32611, USA\\E-mail: klauder@phys.ufl.edu}

\begin{document}
\maketitle
\abstract{
Coherent states can be used for diverse applications in quantum 
physics including the construction of coherent state path integrals. 
Most definitions make use of a lattice regularization; however, recent 
definitions employ a continuous-time regularization that may involve a 
Wiener measure concentrated on continuous phase space paths. The 
introduction of constraints is both natural and economical in coherent 
state path integrals involving only the dynamical and Lagrange 
multiplier variables. A preliminary indication of how these procedures 
may possibly be applied to quantum gravity is briefly discussed. }

\section{Introduction}
Formal expressions for path integrals have been used on a regular basis. 
While mathematically challenged, such formulas are exceedingly useful 
for heuristic insight. Behind every formal expression there should 
stand some well-defined regularization procedure, and we assume that 
to be the case.
In what follows we generally choose units in which $\hbar=1$.

In a familiar notation, the first path integral \cite{fey} was given 
50 years ago by  
  $$\<x''|\,e^{-i\H T}\,|x'\>={\cal N}\int\exp\{i\tint[
\half{\dot x}^2-V(x)]\,dt\}\,\D x\;, $$
and it was soon followed by the more general phase space form 
\cite{fez}
  $$\<x''|\,e^{-i\H T}\,|x'\>={\cal M}\int\exp\{i\tint[p\,{\dot x}-
H(p,x)]\,dt\}\,\D p\,\D x\;,$$
which applies to a wider class of problems.
An alternative phase space path integral based on coherent states 
appeared subsequently \cite{jk1}
  $$\<p'',q''|\,e^{-i\H T}\,|p',q'\>={\cal M}\int\exp\{i\tint[
p\,{\dot q}-H(p,q)]\,dt\}\,\D p\,\D q\;, $$
which has an identical formal definition (but a different 
regularized form). Here, $H(p,q)=\<p,q|\H(P,Q)|p,q\>$ denotes the 
upper symbol for $\H$. A different rule of construction led to another 
coherent state path integral \cite{lie}
  $$\<p'',q''|\,e^{-i\H T}\,|p',q'\>={\cal M}\int\exp\{i\tint[p\,
{\dot q}-h(p,q)]\,dt\}\,\D p\,\D q\;, $$
which involves a different symbol $h(p,q)$ (defined below) for $\H$.

In this brief review we focus on the contribution to path integration 
offered by coherent states. Let us start with the canonical coherent states.

\section{Coherent States}
The canonical coherent states $|p,q\>$ introduced above are defined by
$$|p,q\>\equiv \exp(-iqP)\exp(ipQ) \,|\eta\>\;,$$ 
where $Q$ and $P$ denote an
irreducible pair of self-adjoint operators that satisfy $[Q,P]=i$, along 
with a normalized fiducial vector $|\eta\>$. Such states
are continuously parametrized and admit a resolution of unity in the form
  $$\one =\int |p,q\>\<p,q|\,dm(p,q)\;,\hskip1cm dm(p,q)=dp\,dq/2\pi\;, $$
integrated over $\ir^2$,
for any choice of $|\eta\>$. This relation leads to a Hilbert space 
representation $\psi(p,q)\equiv\<p,q|\psi\>=\<\eta|e^{-ipQ}e^{iqP}|\psi\>$ 
by bounded, continuous functions $\psi$. Each function is a vector in a 
reproducing kernel Hilbert space with reproducing kernel $\<p'',q''|p',q'\>$.

Operators can be defined to act in two different ways. Operators that 
act on the left fulfill
  $$\P'(-p-i\d/\d q,i\d/\d p)\<p,q|\psi\>\equiv\<\eta|\P(P,Q)^\dagger
\exp(-ipQ)\exp(iqP)|\psi\>\;.   $$
If $\P|\eta\>=0$, then $\P$ defines a ``polarization'', which can be 
illustrated by the choice $\P=iP+Q$; in turn, this choice implies that  
$ 0=\<x|[iP+Q]|\eta\>=[\d/\d x+x]\<x|\eta\>  $,
with a solution $\<x|\eta\>\propto \exp(-x^2/2)$.  If $\P|\eta\>=0$, 
then $\P^\dagger\P|\eta\>=0$ and conversely. Indeed, $\<\eta|\P^\dagger\P
|\eta\>=\|\P|\eta\>\|^2=0$ implies that $\P|\eta\>=0$. 

Let us consider the expression 
\bea
  &&\hskip-.5cm K^\nu(p,q;p',q')\equiv\exp[-(\nu/2) T (\P'^\dagger\P')
(-p-i\d/\d q,i\d/\d p)]\no\\
&&\hskip3.3cm\times\,\delta(p-p')\,\delta(q-q')\;,\no
\eea
where $(\P'^\dagger\P')=[-iP+Q][iP+Q]\equiv(p+i\d/\d q)^2+(i\d/\d p)^2-1$,
and  reexpress it in a conventional (not coherent state!) phase space 
path integral as
\bea
 &&\hskip-.9cm  K^\nu(p'',q'';p',q')\no\\
&& ={\cal M}\int\exp\{i\tint[k{\dot q}-x{\dot p}]\,dt\no\\
&&\hskip1cm-(\nu/2)\tint[(k-p)^2+x^2-1]\,dt\}\,\D k\,\D x\,\D p\,\D q\no\\
 &&={\cal M}\int\exp\{i\tint[p\,{\dot q}+k\,{\dot q}-x\,{\dot p}]\,dt\no\\
&&\hskip1cm-(\nu/2)\tint[k^2+x^2-1]\,dt\}\,\D k\,\D x\,\D p\,\D q\;.
\no\eea
Carrying out the integrations over $k$ and $x$ leads to
\bea
 &&\hskip-.8cmK^\nu(p'',q'';p',q')={\cal N}\int e^{i\tint p\,{\dot q}\,dt}
\,e^{-(1/2\nu)\tint({\dot p}^2+{\dot q}^2)\,dt}\,\,\D p\,\D q \no\\
&& \hskip1.8cm =2\pi e^{\nu T/2}\int e^{i\tint p\,dq}\,d\mu^\nu_W(p,q)\;,\no
\eea
where we define the stochastic integral $\tint p\,dq$ by the midpoint 
(Stratonovich) rule.
This final expression exhibits a well-defined path integral over Wiener 
measure without any ambiguity whatsoever. 

The limit of $K^\nu$ as $\nu\ra\infty$ yields an integral kernel for the 
projection operator onto the subspace for which $\P'\psi(p,q)=0$, i.e., 
the subspace of interest. Hence, we conclude that
\bea
  &&\hskip-1cm\<p'',q''|p',q'\>=\exp\{i\half(p''+p')(q''-q')-
\quarter[(p''-p')^2+(q''-q')^2]\}\no\\
  &&\hskip1.03cm=\lim_{\nu\ra\infty}2\pi e^{\nu T/2}\int e^{i\tint p\,dq}
\,d\mu^\nu_W(p,q)\;. \no
\eea
This a very significant formula.

In evaluating any conditionally convergent integral, such as $\tint 
e^{-iy^2}\,dy$, $-\infty<y<\infty$, it is first necessary to define the 
formal expression. This may be done, for example, by adopting $\tint 
e^{-iy^2 -\epsilon y^2}\,dy$ in the limit that $\epsilon\ra 0^+$, which 
then gives one possible definition. The path integral regularization 
given above is fundamentally the same as the one-dimensional example, and 
it may be given the following interpretation: In defining the coherent 
state overlap, we have introduced a Brownian motion regularization into the 
ill-defined, conditionally convergent (at best!)  path integral,
  $${\cal M}\int e^{i\tint p\,{\dot q}\,dt}\,\D p\,\D q \;,$$
and instead defined this formal expression by the preceding regularized 
expression, removing the regularization (i.e., $\nu\ra\infty$) as a final 
step. Evidently, the only price to pay to ensure a well-defined path 
integral expression is the adoption of a {\it metric on phase space} to 
support the Brownian motion, which in the present case corresponds to a 
flat space expressed in Cartesian coordinates. The resultant quantization 
{\it automatically} leads to a canonical coherent state representation, 
which is entirely equivalent to choosing the polarization $\P=Q+iP$. 

It is noteworthy that Brownian  motion regularization on other geometries, 
e.g., a sphere or a pseudosphere, leads to alternative quantizations 
when $\nu\ra\infty$, namely, quantization in terms of the generators of 
SU(2) or SU(1,1), respectively.~\cite{dk1,dkp,jk2} 

\section{Dynamics}
The second form of introducing operators is those that act on the right. 
Observe for any $|\psi\>$ and $|\eta\>$ that
  $$\H(-i\d/\d q,q+i\d/\d p)\,\<p,q|\psi\>=\<\eta|\exp(-ipQ)\exp(iqP)
\,\H(P,Q)\,|\psi\>\;.$$
In canonical coherent states, therefore, Schr\"odinger's equation assumes 
the form
  $$i\,\d\,\psi(p,q,t)/\d t=\H(-i\d/\d q,q+i\d/\d p)\,\psi(p,q,t)\;. $$

The solution to Schr\"odinger's equation has a coherent state path 
integral given by~\cite{dk1} 
%\bea
 $$\<p'',q''|\,e^{-i\H T}\,|p',q'\>
 =\lim_{\nu\ra\infty}2\pi e^{\nu T/2}\int e^{i\tint[p\,dq-h(p,q)\,dt]}
\,d\mu^\nu_W(p,q)\;, $$
%\eea
where $h(p,q)$ denotes the lower symbol for the quantum Hamiltonian, i.e.,
  $$\H(P,Q)=\tint h(p,q)\,|p,q\>\<p,q|\,dp\,dq/2\pi\;. $$

After a canonical coordinate transformation classically generated by 
${\o p}\,d{\o q}=p\,dq+dF({\o q},q)$ and based on the midpoint rule for 
which the ordinary rules of calculus apply, even for Brownian paths, we 
determine that
  $$\<{\o p}'',{\o q}''|\,e^{-i\H T}\,|{\o p}',{\o q}'\>
 =\lim_{\nu\ra\infty}2\pi e^{\nu T/2}\int e^{i\tint[{\o p}\,d{\o q}+
d{\o G}({\o p},{\o q})-{\o h}({\o p},{\o q})\,dt]}\,d
{\o \mu}^\nu_W({\o p},{\o q})\;, $$
for some function ${\o G}({\o p},{\o q})$ and where 
$|{\o p},{\o q}\>\equiv|p({\o p},{\o q}), q({\o p},{\o q})\>=
|p,q\>$, ${\o h}({\o p},{\o q})=h(p({\o p},{\o q}), q({\o p},{\o q}))
=h(p,q)$, and $d{\o \mu}^\nu_W$ is Weiner measure on the plane expressed, 
generally, in curvilinear coordinates. Moreover,
 $$\H(P,Q)=\tint {\o h}({\o p},{\o q})\,|{\o p},{\o q}\>\<{\o p},{\o q}|
\,d{\o p}\,d{\o q}/2\pi\;.  $$

Observe, therefore, that the path integral and resultant propagator are 
{\it covariant} under general canonical coordinate transformations. This 
fact gives rise to the assertion that we deal here with a coordinate-free 
formulation of quantization.
\section{Metrical Quantization}
The preceding analysis suggests a novel two-step quantization procedure. 
First, add a metric to the classical phase space that can be used to keep 
track of  the physical meaning of mathematical expressions under coordinate 
transformations.~\cite{jk3} For a flat space expressed in Cartesian 
coordinates, $d\sigma^2=dp^2+dq^2$---and in these coordinates 
$\half(p^2+q^2)+q^4$ physically corresponds to a quartic anharmonic 
oscillator---or, for a space of constant negative curvature, 
$d\sigma^2=\beta^{-1}q^2\,dp^2+\beta q^{-2}\,dq^2$, $q>0$, $\beta$ a 
constant, etc. Second, use the metric in a Wiener measure regularization 
of an otherwise formal  phase space path integral. For example, for the 
negative curvature case with $\beta>1/2$,
\bea 
 &&K\equiv\lim_{\nu\ra\infty}{\cal N}\int e^{i\tint[p\,{\dot q}-h(p,q)]
\,dt}\,e^{-(1/2\nu)\tint[\beta^{-1}q^2{\dot p}^2+\beta q^{-2}{\dot q}^2]
\,dt}\,\D p\,\D q\no\\
&& \hskip.45cm  =\lim_{\nu\ra\infty}2\pi[1-1/2\beta]e^{\nu T/2}\int 
e^{i\tint[p\,dq-h(p,q)\,dt]}\,d\tau^\nu_W(p,q)\;.\no
\eea
This expression leads to a positive-definite function, and that fact 
alone allows one to conclude that
\bea
  &&K=\<p'',q''|\,e^{-i\H T}\,|p',q'\>\;,\hskip1cm |p,q\>\equiv e^{-ipq}
e^{ipQ}e^{-i(\ln q)D}\,|\beta\>\;,\no\\
&&\hskip1cm [Q,D]=iQ,\hskip.5cm Q>0,\hskip.5cm (Q-1+i\beta^{-1} D)|\beta\>
=0\;,\no\\
  &&\hskip1.4cm \H=\int h(p,q)\,|p,q\>\<p,q|\,dp\,dq/2\pi[1-1/2\beta]\;, \no\\
&&\hskip2cm \one=\int|p,q\>\<p,q|\,dp\,dq/2\pi[1-1/2\beta]\;.\no
 \eea

\section{Constraints}
General constraints may be imposed in the quantum theory by introducing a 
projection operator $\E$ from the original Hilbert space $\H$ onto the 
physical Hilbert space $\H_{{\rm phys}}\equiv\E\H$. If $\Phi_\alpha(P,Q)$, 
$\alpha=1,\dots,A$, denote hermitian quantum constraints, it may be argued 
that it suffices to choose 
  $$\E=\E(\!\!(\Phi_\alpha\,M^{\alpha\,\beta}\Phi_\beta\le\delta(\hbar)^2)
\!\!) $$
where $\{M^{\alpha\,\beta}\}$ is a symmetric, positive-definite matrix, and 
$\delta(\hbar)$ is chosen so that $\E$ projects onto a suitably small space. 
By choosing an appropriate measure for the Lagrange multiplier variables, 
it is possible to generate the expression
  $$\<p'',q''|\E e^{-i(\E\H\E)T}\E|p',q'\> $$
corresponding to  temporal evolution entirely within the physical subspace.

In this generality, and by using appropriate integration measures for the 
Lagrange multiplier variables along with the usual measure for the dynamical
variables, coherent state path integrals may be constructed to deal with a
general set of constraints.~\cite{kk3}

\section{Preliminary Application to Gravity}
The classical phase space variables for $3+1$ gravity are symmetric 
$3\times 3$ matrices $\{\pi^{jk}\}$ and $\{g_{jk}\}>0$ with the latter 
being positive definite. A natural metric on phase space is given by
  $$d\sigma^2=\tint_\Sigma\,[g^{-1/2}g_{jk}\,g_{lm}\,d\pi^{kl}\,
d\pi^{jm}+g^{1/2}g^{jk}\,g^{lm}\,dg_{kl}\,dg_{jm}]\,d^3\!x\;,$$
an expression which is diffeomorphism invariant under coordinate 
transformations on the space like surface $\Sigma$. Adopting the 
viewpoint of metrical quantization, we propose that
\bea &&\<\pi'',g''|\pi',g'\>\equiv\lim_{\nu\ra\infty}{\cal N}\int 
e^{i\tint\pi^{jk}\,{\dot g}_{jk}\,d^3\!x\,dt}\no\\
&&\hskip1cm\times e^{-(1/2\nu)\tint[g^{-1/2}g_{jk}g_{lm}\,
{\dot \pi}^{kl}{\dot \pi}^{jm}+g^{1/2}g^{jk}g^{lm}\,{\dot g}_{kl}
{\dot g}_{jm}]\,d^3\!x\,dt}\,\Pi_{a\le b}\,\D\pi^{ab}\,\D g_{ab} \no
\eea
fully determines a reproducing kernel for gravity {\it before} any 
constraints are imposed. It is noteworthy that this formal functional 
integral is {\it ultralocal}, i.e., has no spatial derivatives. 
Recent advances in the formulation and solution of ultralocal quantum 
field theories suggest that evaluation of this reproducing kernel may 
indeed be possible.~\cite{kkk}

%\section*{References}

\end{document}